\documentstyle[prl,aps,preprint]{revtex}

\begin{document}
\draft

\title{Geometric phases and anholonomy for
a class of chaotic classical systems}
\author{C. Jarzynski}
\address{University of California, Department of Physics,\\
and Lawrence Berkeley Laboratory\\
Berkeley, CA~~94720\\
{\rm beginning July, 1994:}\\
Institute for Nuclear Theory, University of Washington,\\
Seattle, WA~~98195}
\date{\today}
\maketitle

\begin{abstract}
Berry's phase may be viewed as
arising from the parallel transport of
a quantal state around a loop in parameter space.
In this Letter, the classical limit of this
transport is obtained
for a particular class of chaotic systems.
It is shown that this ``classical parallel transport''
is anholonomic --- transport around a closed curve
in parameter space does not bring a point in
phase space back to itself ---
and is intimately related to the
Robbins-Berry classical two-form.
\end{abstract}


Ordinarily, {\it Berry's phase} is defined as
the unexpected phase $\Delta\gamma_n$
picked up by a quantal eigenstate
$\vert n\rangle$ evolving under a parameter-dependent
Hamiltonian $\hat h({\bf R})$, when ${\bf R}$ is made
to slowly trace out a loop $\Gamma$ in parameter
space\cite{berry,breview}.
This phase is {\it geometric}:
its value is given in terms of the flux
of a 2-form ${\bf V}_n({\bf R})=-i\hbar\langle\nabla n\vert\times
\vert\nabla n\rangle$ through a surface bounded by
$\Gamma$.
(Here, $\vert n\rangle$
is the $n$th eigenstate of $\hat h$, and
$\vert\nabla n\rangle\equiv
\partial\vert n\rangle/\partial{\bf R}$.
{\bf R}-space is taken to be ordinary 3-space, hence
2-forms are simply vector fields.)
The search for classical counterparts of Berry's phase
has been particularly challenging for chaotic systems.
While Robbins and Berry\cite{rb} have obtained
the classical limit of the 2-form
${\bf V}_n({\bf R})$, the analogue of the phase
$\Delta\gamma_n$ itself has proven elusive.
It is the purpose here to show that,
if one interprets Berry's phase as an {\it anholonomy}
effect\cite{breview,simon}, rather than as the
dynamical effect mentioned above,
then for a certain class of chaotic
systems, $\Delta\gamma_n$ does indeed have a
classical analogue.

This Letter is arranged as follows.
First, the interpretation of Berry's phase as
an anholonomy effect --- arising from the parallel
transport of an eigenstate around a loop in parameter
space --- is reviewed.
Next, this transport is expressed in terms
of its generator $\hat{\bf\xi}({\bf R})$.
The classical limit ${\bf\xi}(z,{\bf R})$ of
this operator is then obtained, and {\it classical
parallel transport} is defined as the flow in
phase space ($z$-space) generated by ${\bf\xi}(z,{\bf R})$.
Finally, this flow is studied, and the classical
analogue of $\Delta\gamma_n$ is derived.
Results similar to those presented here have
been obtained independently by Dr. Jonathan
Robbins (personal communication).

First, let us define
$h(z,{\bf R})$ (or simply $h({\bf R})$) to be the
classical Hamiltonian which
is the classical limit of $\hat h({\bf R})$.
Motion under $h({\bf R})$,
with {\bf R} fixed, is assumed bounded, and ergodic
over the energy shell.
(This implies that, under the Poisson bracket,
$h(z,{\bf R})$ commutes only with functions of
the form $f(h,{\bf R})$, a fact which will come
in handy.)
Phase space is $2N$-dimensional, where $N>1$.

Now, consider the following definition of the
``parallel transport''
of a quantal state in {\bf R}-space:
along a curve ${\bf R}(\tau)$ starting from
${\bf R}_0$, the
$n$th eigenstate of $\hat h({\bf R}_0)$ gets transported
to the $n$th eigenstate of $\hat h({\bf R}^\prime)$,
for each point ${\bf R}^\prime$ along the curve.
To remove ambiguity about how the {\it phase}
of the state changes along the curve, we impose the condition
$\langle\psi(\tau)\vert\psi(\tau+\delta\tau)\rangle
=1+{\it O}((\delta\tau)^2)$ on an eigenstate
thus transported.
(Here, $\tau$ is a dummy variable,
labelling points along the curve in ${\bf R}$-space,
and also labelling the state $\vert\psi\rangle$ found
at each such point.)
If we extend this definition to include linear
combinations of eigenstates (by assuming the principle
of superposition)
then we have a prescription for how an
arbitrary state $\vert\psi\rangle$ ``evolves''
along an arbitrary curve in parameter space.
It was shown by Simon\cite{simon} that, under
the transport thus defined,
an eigenstate $\vert n\rangle$ taken around a
loop $\Gamma$ picks up a net phase equal to Berry's phase.
Berry's phase thus emerges as an {\it anholonomy}
effect: when {\bf R} completes its
circuit, the quantal phase does not
return to its initial value.
(The term ``anholonomy'' refers to
the situation when a non-zero change in
some quantity is induced by taking a parameter
{\bf R} around a closed circuit\cite{breview}.
For instance, a vector parallel transported around
a loop in a curved space typically
does not return to its original orientation.)

The quantal parallel transport thus defined is
unitary, and may be described in terms of its
{\it generator}, the vector operator
$\hat{\bf\xi}({\bf R})$ such that
\begin{equation}
\label{qgen}
i\hbar{d\over d\tau}\vert\psi\rangle
\,=\,{d{\bf R}\over d\tau}
\cdot\hat{\bf\xi}\,\vert\psi\rangle
\end{equation}
for any $\vert\psi(\tau)\rangle$ enjoying
parallel transport along ${\bf R}(\tau)$.
(I.e. if  parallel transport is a rule
for associating an infinitesimal step $\vert\delta\psi\rangle$
in Hilbert space with a given step $\delta{\bf R}$
in parameter space, then $\hat{\bf\xi}({\bf R})$ is
the operator such that $i\hbar\vert\delta\psi\rangle=
\delta{\bf R}\cdot\hat{\bf\xi}\vert\psi\rangle$.)
It is a straightforward exercise to show that $\hat{\bf\xi}$
is determined by the conditions
\begin{eqnarray}
\label{qcom}
[\hat{\bf\xi},\hat h]\,&=&\,i\hbar
(\nabla\hat h-\hat{\bf D})\\
\label{qzero}
\langle n\vert\hat{\bf\xi}\vert n\rangle\,&=&\,0,
\end{eqnarray}
where $\langle m\vert\hat{\bf D}\vert n\rangle
\equiv\langle m\vert\nabla\hat h\vert n\rangle
\,\delta_{mn}$.
(Eqs.\ref{qcom} and \ref{qzero}, and their
classical counterparts \ref{com} and \ref{zero},
appear as well
in the context of Born-Oppenheimer forces, where {\bf R}
is a dynamical quantity rather than an externally
driven parameter; see Aharonov et al\cite{aharonov}.)

Given this quantal picture, what might the
corresponding classical picture be?
Since Eq.\ref{qgen} is essentially
a prescription for lifting
a curve ${\bf R}(\tau)$ from parameter space
to a curve $\vert\psi(\tau)\rangle$ in
Hilbert space (given an initial state
$\vert \psi_0\rangle$), we might expect
the classical version to be a prescription for
lifting ${\bf R}(\tau)$ to a curve $z(\tau)$ in
{\it phase space} (given an initial $z_i$).
With this in mind,
let us obtain the vector function ${\bf\xi}(z,{\bf R})$
which is the classical limit of $\hat{\bf\xi}({\bf R})$,
and then define {\it classical
parallel transport} to be the flow in phase space
generated by ${\bf\xi}(z,{\bf R})$, according to
\begin{equation}
\label{xiflow}
{dz\over d\tau}\,=\,
{d{\bf R}\over d\tau}\cdot
\{z,{\bf\xi}\}.
\end{equation}
(As in the quantal, case, $d{\bf R}/d\tau\cdot
\hat{\bf\xi}$ acts as a Hamiltonian along
${\bf R}(\tau)$.
This makes classical parallel transport a
canonical flow, just as the quantal version
is unitary.)
Once $\bf\xi$ is obtained, we will be able to
study what happens when a point $z_i$ gets
transported around a closed curve in parameter
space:
will $z_i$ return to itself, and if not will the
anholonomy bear any resemblance to Berry's phase?

To obtain ${\bf\xi}(z,{\bf R})$ from $\hat{\bf\xi}({\bf R})$,
we exploit the fact that $\hat{\bf\xi}({\bf R})$ is
specified by Eqs.~\ref{qcom} and \ref{qzero},
whose classical limits, using the simplest of
semiclassical approximations, are:
\begin{eqnarray}
\label{com}
\{ {\bf\xi},h\}\,&=&\,\nabla h-\langle\nabla h
\rangle_{h,{\bf R}}\\
\label{zero}
\langle{\bf\xi}\rangle_{E,{\bf R}}\,&=&\,0.
\end{eqnarray}
Here, $\langle\cdots\rangle_{E,{\bf R}}$ denotes
a phase space average over an energy shell of $h({\bf R})$:
\begin{equation}
\langle\cdots\rangle_{E,{\bf R}}\,=\,
\Bigl({\partial\Omega\over\partial E}\Bigr)
^{-1}\,
\int dz\,\delta(E-h)\,\cdots\,
\end{equation}
where $\Omega(E,{\bf R})=\int dz\,\theta(E-h)$
is the volume of phase space enclosed by this
shell.
Note that the left side of Eq.\ref{com}
is evaluated at some $(z,{\bf R})$;
the value of the subscript $h$ on the right
side is evaluated at the same $(z,{\bf R})$.

Eqs.~\ref{com} and \ref{zero} are solved by:
\begin{equation}
\label{soln}
{\bf\xi}(z,{\bf R})\,=\,
\lim_{\alpha\rightarrow 0}\,
\int_{-\infty}^0 ds\,
{\rm e}^{-\alpha\vert s\vert}
\,\nabla\tilde h(z_s,{\bf R}),
\end{equation}
where, as in Ref.~\onlinecite{br},
$\nabla\tilde h\equiv\nabla h-\langle\nabla h\rangle_{h,{\bf
R}}$,
and $z_s(z,{\bf R})$ is the point in phase space reached by
evolution from $z$, for time $s$, under $h({\bf R})$.
If Eq.~\ref{soln} converges, it solves
Eqs.~\ref{com} and \ref{zero} uniquely.
Generically, however, Eq.~\ref{soln} approaches an
infinitely convoluted {\it distribution}\cite{br}, rather
than a smooth vector function on phase space.
Since it is not clear how such an object
could serve as the generator of flow, we
assume that Eq.~\ref{soln} converges.
This places a restriction (discussed below)
on the class of systems for which
the results to be derived are valid.

Eqs.\ref{xiflow} and \ref{soln} together specify
how a point in phase space gets transported
along a curve in parameter space.
This flow has a number of properties that
simplify the investigation of transport around
a {\it closed} curve:

$\bullet$ First, consider a new Hamiltonian
$h^\prime$ of the form
$h^\prime(z,{\bf R})\,=\,
f\Bigl(h(z,{\bf R}),{\bf R}\Bigr)$.
Then any $\bf\xi$ satisfying Eq.~\ref{com}
also satisfies
$\{{\bf\xi},h^\prime\}=\nabla h^\prime-
\langle\nabla h^\prime\rangle$,
i.e. $h({\bf R})$ and $h^\prime({\bf R})$
give rise to the same generator $\bf\xi({\bf R})$.
Since $h^\prime({\bf R})$ simply relabels
the energy shells of $h({\bf R})$,
this means that {\it parallel transport
is determined by the manner in which $h({\bf R})$
divides phase space into energy shells} (at each ${\bf R}$),
but is independent of the energy values that happen to be
assigned to those shells.
(In Ref.\cite{rb}, where the classical limit of the
quantal 2-form ${\bf V}_n$ is obtained,
it is shown that the resulting {\it classical
2-form} ${\bf V}^c$ shares this property.)

Thus, for studying the flow generated by ${\bf\xi}({\bf R})$,
we may replace $h({\bf R})$ with any $h^\prime({\bf R})$
of the form given above.
A convenient replacement is the ``volume
Hamiltonian''\cite{rb}
\begin{equation}
\Omega_P(z,{\bf R})\,=\,
\Omega\Bigl(h(z,{\bf R}),{\bf R}\Bigr),
\end{equation}
with $\Omega$ as defined earlier.
$\Omega_P({\bf R})$ relabels the energy shells
of $h({\bf R})$, assigning to each a value equal to
the volume of phase space it encloses.
Note that the dynamics under $\Omega_P(z,{\bf R})$,
with ${\bf R}$ fixed, is the same as under
$h(z,{\bf R})$, only with time re-scaled.

$\bullet$ For an arbitrary curve ${\bf R}(\tau)$ from
${\bf R}_0$ to ${\bf R}_1$, parallel transport
defines a mapping $z\rightarrow z^\prime$ of
phase space onto itself:
$z^\prime$ is the point reached by transport
from $z$ along ${\bf R}(\tau)$.
Let $z(\tau)$ then denote the curve connecting
$z$ to $z^\prime$, satisfying Eq.\ref{xiflow}.
{}From Eq.\ref{com}, and the identity
$\langle\nabla\Omega_P\rangle_{E,{\bf R}}=0$,
we get
\begin{equation}
\label{conserv}
{d\over d\tau}\,\Omega_P\Bigl(
z(\tau),{\bf R}(\tau)\Bigr)\,=\,0.
\end{equation}
In words, {\it parallel transport conserves the value of}
$\Omega_P(z,{\bf R})$
(just as the quantal version conserves quantum number $n$.)
This tells us that the canonical transformation
$z\rightarrow z^\prime$
maps each energy shell $A_0$ of $\Omega_P({\bf R}_0)$
to the same-valued shell $A_1$ of $\Omega_P({\bf R}_1)$.
(Thus, under parallel transport around a {\it closed}
curve, $A_0$ gets mapped to itself.)

$\bullet$ While the statement that parallel transport takes
$A_0$ to $A_1$ is true for any ${\bf R}(\tau)$ from
${\bf R}_0$ to ${\bf R}_1$, the specific point $z^\prime$
on $A_1$ to which a given $z$ on $A_0$ gets mapped depends
on ${\bf R}(\tau)$.
Thus, a slight change $\delta{\bf R}(\tau)$ in the
path from ${\bf R}_0$ to ${\bf R}_1$ induces a
slight shift $\delta z^\prime$ in the final point $z^\prime$.
The mapping $z^\prime\rightarrow z^\prime+
\delta z^\prime$
induced by this change of path is itself a canonical
transformation which maps $A_1$ onto itself.
Since any continuous deformation of ${\bf R}(\tau)$
(with endpoints fixed)
may be constructed from a sequence of such
infinitesimal changes in path, we are led to conclude
that, {\it by continuously deforming the path from
${\bf R}_0$ to ${\bf R}_1$, we generate a canonical
flow of the point $z^\prime$ to which a fixed
$z$ gets mapped}.
Now, the assumption of ergodicity
guarantees that the only canonical flow on
the energy shell $A_1$ is that generated by
$\Omega_P(z,{\bf R}_1)$ --- or $h(z,{\bf R}_1)$ --- itself.
Thus, a continuous deformation of the path from
${\bf R}_0$ to ${\bf R}_1$ displaces $z^\prime$
along a trajectory of the Hamiltonian $\Omega_P({\bf R}_1)$.
Assuming further that any {\it closed}
curve in {\bf R}-space
can be shrunk to a single point, we
conclude that, if a point $z_i$ is transported around
a curve starting and ending at ${\bf R}_0$, to
a point $z_f$, then $z_i$ and $z_f$ lie on a single
trajectory of $h({\bf R}_0)$.

Since the energy shells of $h({\bf R})$ are identical
to those of $\Omega_P({\bf R})$, parallel transport from
${\bf R}_0$ to ${\bf R}_1$ maps the
shells of $h({\bf R}_0)$ to those of $h({\bf R}_1)$.
This mapping conserves the value of $\Omega_P$,
but not necessarily of $h$.
Thus, the assumption that Eq.\ref{soln} converges
has restricted us to systems with
the non-generic property that (for any ${\bf R}_0$
and ${\bf R}_1$), $h(z,{\bf R}_0)$ is related
to $h(z,{\bf R}_1)$ by a canonical
transformation, along with a possible relabelling
of the values of the energy shells.
Let us call such a system a {\it generalized canonical
family}.
This follows Ref.\cite{robbins},
where a {\it canonical family} is defined by the
property that the Hamiltonians at different
points in {\bf R}-space are related by
canonical transformation, without any relabelling
of energy shells.
$\Omega_P(z,{\bf R})$ constitutes a canonical family.

As discussed in  Ref.\cite{robbins},
for a canonical family, e.g. $\Omega_P(z,{\bf R})$,
we may construct a vector function
${\bf g}(z,{\bf R})$ so that the flow generated
by ${\bf g}$ (as per Eq.\ref{xiflow}, only
with ${\bf g}$ in place of ${\bf\xi}$)
has the following two properties.
First, as with ${\bf\xi}$, this flow,
along any path from ${\bf R}_0$ to ${\bf R}_1$,
maps an energy shell of $\Omega_P({\bf R}_0)$
to the same-valued shell of $\Omega_P({\bf R}_1)$.
However, unlike with ${\bf\xi}$,
this mapping is {\it independent} of
the path connecting ${\bf R}_0$ to ${\bf R}_1$.
The generator ${\bf g}$ provides the final tool
needed to establish exactly where a point
$z_i$ gets taken when parallel transported around a
closed loop in parameter space.

Since flow under {\bf g} preserves $\Omega_P$,
we have $\{{\bf g},\Omega_P\}=\nabla\Omega_P$,
as was the case with $\bf\xi$.
Thus, $\{{\bf\xi}-{\bf g},\Omega_P\}=0$, so
{\bf g} and $\bf\xi$ differ at most by some function
${\bf A}(\Omega_P,{\bf R})$:
${\bf\xi}(z,{\bf R})\,=\,
{\bf g}(z,{\bf R})\,+\,
{\bf A}\Bigl(\Omega_P(z,{\bf R}),{\bf R}\Bigr)$.
Since the phase space average of ${\bf\xi}$ over
any energy shell of $\Omega_P({\bf R})$ is zero,
we have
\begin{equation}
{\bf A}(\omega,{\bf R})\,=\,
-\langle{\bf g}(z,{\bf R})\rangle_{\omega,{\bf R}},
\end{equation}
where $\omega$ denotes the volume
enclosed by the energy shell of $\Omega_P({\bf R})$
over which the average is taken.
As shown in Ref.\onlinecite{robbins},
we may choose {\bf g} so that
\begin{equation}
\label{curlav}
\nabla\times{\bf A}(\omega,{\bf R})\,=\,
{\bf V}^c(\omega,{\bf R}),
\end{equation}
where ${\bf V}^c$ is the Robbins-Berry
classical 2-form (the classical limit of the
quantal 2-form ${\bf V}_n$) associated with
$\Omega_P(z,{\bf R})$.
In what follows we assume Eq.~\ref{curlav} holds.

We are finally prepared to investigate parallel
transport around a loop $\Gamma$,
starting and ending at ${\bf R}_0$.
Let ${\bf R}(\tau)$ explicitly represent this loop, with
$\tau$ running from $\tau_i$ to $\tau_f$.
Let $z_i$ be an initial point in phase space, and
$z_f$ the point reached from $z_i$ by
transport around $\Gamma$.
We already know that $z_f$ and $z_i$ lie
on a single trajectory of $\Omega_P({\bf R}_0)$;
let us use $\Delta\sigma$, the time of evolution under
$\Omega_P({\bf R}_0)$ separating $z_f$ from $z_i$,
as the measure of ``distance'' between these two points
along the trajectory.
$\Delta\sigma$ thus measures
the anholonomy associated with transport around $\Gamma$.
(This definition of distance along a trajectory
is meant to
be unaffected by a relabelling of the energy shells:
for {\it any} Hamiltonian $h$, we take distance along a
trajectory to mean time of evolution
under the associated ``volume Hamiltonian''
$\Omega_P$.
In one degree of freedom, this reduces
to the ordinary angle variable of action-angle
variables, divided by $2\pi$.)

To solve for $\Delta\sigma$, consider the mapping
$z\rightarrow y(z,{\bf R})$, where $y$
is reached from $z$ by flow under {\bf g}
along any path {\it from}
${\bf R}$ {\it to} ${\bf R}_0$.
This constitutes a kind of projection:
for any {\bf R}, each $z$ on a given energy shell
of $\Omega_P({\bf R})$ is mapped to a point $y$
on the corresponding shell of $\Omega_P({\bf R}_0)$.
$y(z,{\bf R})$ satisfies
\begin{equation}
\label{neg}
\nabla y(z,{\bf R})\,=\,
-\Bigl\{
y(z,{\bf R}),{\bf g}(z,{\bf R})\Bigr\}.
\end{equation}

Now, let $z(\tau)$ be the phase space curve obtained by
parallel transport from $z_i$ along ${\bf R}(\tau)$,
and let $y(\tau)\equiv y(z(\tau),{\bf R}(\tau))$.
Thus, as $z(\tau)$ traces out some path in phase
space, starting and ending at an energy shell $A_0$ of
$\Omega_P({\bf R}_0)$, its projection $y(\tau)$
traces out a path wholly confined to $A_0$.
Since ${\bf R}(\tau_i)={\bf R}(\tau_f)={\bf R}_0$,
we have $y(\tau_i)=z_i$, and
$y(\tau_f)=z_f$, so we solve for the distance
between $z_i$ and $z_f$
by solving for $y(\tau)$:
\begin{eqnarray}
{dy\over d\tau}\,&=&\,
{\partial y\over\partial z}\,{dz\over d\tau}
\,+\,\nabla y\cdot{d{\bf R}\over d\tau}\nonumber\\
&=&\,
{\partial y\over\partial z}
{d{\bf R}\over d\tau}\cdot\Bigl\{
z,{\bf\xi}\Bigr\}\,-\,
{d{\bf R}\over d\tau}\Bigl\{
y,{\bf g}\Bigr\}\nonumber\\
&=&\,
{d{\bf R}\over d\tau}\cdot\Bigl\{
y,{\bf A}\Bigr\},
\end{eqnarray}
using Eqs.~\ref{xiflow} and \ref{neg},
along with properties of the Poisson bracket.
In the last line,
${\bf A}={\bf A}(\Omega_P(z,{\bf R}),{\bf R})
={\bf A}(\Omega_P(y,{\bf R}_0),{\bf R})$, hence
\begin{equation}
{dy\over d\tau}\,=\,
{d{\bf R}\over d\tau}\cdot
{\partial{\bf A}\over\partial\omega}
(\omega,{\bf R})\cdot
\Bigl\{y,\Omega_P(y,{\bf R}_0)\Bigr\},
\end{equation}
with $\partial{\bf A}/\partial\omega$ evaluated at
the constant value
$\omega=\Omega_P(y,{\bf R}_0)$.
Finally, defining
\begin{equation}
\sigma(\tau)\,=\,
\int_{{\bf R}_0}^{{\bf R}(\tau)} d{\bf R}^\prime\cdot
{\partial{\bf A}\over\partial\omega}(\omega,{\bf R}^\prime),
\end{equation}
with integration occurring along
$\Gamma$, we get
\begin{equation}
{dy\over d\sigma}\,=\,
\Bigl\{y,\Omega_P(y,{\bf R}_0)\Bigr\}.
\end{equation}
This means that $y$ evolves along a trajectory
of $\Omega_P({\bf R}_0)$, with $\sigma$ playing
the role of time of evolution.
Thus, the distance $\Delta\sigma$
separating $z_f=y(\tau_f)$ from
$z_i=y(\tau_i)$ is given by $\sigma(\tau_f)-
\sigma(\tau_i)=\oint_\Gamma d{\bf R}\cdot
\partial{\bf A}/\partial\omega$.
Using Stokes's theorem and Eq.\ref{curlav},
we finally have
\begin{equation}
\label{central}
\Delta\sigma\,=\,
{\partial\over\partial\omega}\,
\int\int d{\bf S}\cdot{\bf V}^c(\omega,{\bf R}),
\end{equation}
where the integral denotes the flux of
${\bf V}^c(\omega,{\bf R})$
through a surface bounded by $\Gamma$.

We have taken ${\bf V}^c(\omega,{\bf R})$ to be the 2-form
associated with $\Omega_P(z,{\bf R})$.
However, since this 2-form is unchanged by a relabelling
of the energy shells\cite{rb}, ${\bf V}^c(\omega,{\bf R})$
is equally well the 2-form associated with our original
$h(z,{\bf R})$.
(In either case, however, the integral
$\int\int d{\bf S}\cdot{\bf V}^c$ is evaluated at fixed
$\omega=\Omega(E,{\bf R})$, not at fixed $E$.)

Eq.~\ref{central} is the central result of this Letter;
it gives the classical anholonomy resulting from parallel
transport around a loop in parameter space.
But, does it make sense to call this anholonomy
the chaotic classical analogue of Berry's phase?
Let us focus on the fact that both
$\Delta\sigma$ and Berry's phase $\Delta\gamma_n$
are expressed geometrically,
in terms of the flux of a 2-form
(${\bf V}^c$ or ${\bf V}_n$) through the loop
in parameter space.
Specifically, since
$\Delta\gamma_n=-(1/\hbar)\int\int d{\bf S}\cdot
{\bf V}_n$\cite{rb},
and since ${\bf V}_n\rightarrow {\bf V}^c$
semiclassically, Eq.\ref{central}
suggests the correspondence
\begin{equation}
\Delta\sigma\,=\,
-\hbar{\partial\over\partial\omega}
\Delta\gamma_n
\end{equation}
between the classical and quantal
measures of anholonomy.
This relationship is the same as that found in Ref.\cite{berry85}
between Berry's phase and {\it Hannay's angle}\cite{hannay}
(the classical geometric phase for integrable
systems),
which suggests that we are on the right track in
associating $\Delta\sigma$ with $\Delta\gamma_n$.

On the other hand, the original formulation of Berry's
phase makes a {\it dynamical} statement concerning
evolution under a slowly time-dependent Hamiltonian.
Does $\Delta\sigma$ have a similar
significance for chaotic classical systems?
That is, does it make a prediction concerning the
evolution of trajectories under $h(z,{\bf R})$, when
{\bf R} slowly traces out a closed curve?
The exponential divergence of chaotic trajectories
makes this a difficult question, to which I
have no answer.
For the time being, then, Eq.\ref{central} is a
purely formal result, one more piece of the
puzzle, but not the last piece.
Its value lies in emphasizing and illuminating
the classical chaotic limit of the quantal
transport underlying Berry's phase,
and in demonstrating that the {\it anholonomy}
associated with this classical transport
closely resembles its quantal counterpart.

As mentioned, the 2-form ${\bf V}^c$ was originally
derived as the classical limit of ${\bf V}_n$.
Later, Berry and Robbins\cite{br} demonstrated
the significance of ${\bf V}^c$ within a purely
classical context.
Namely, when ${\bf R}(t)$ is itself a dynamical quantity,
${\bf V}^c$ acts as a magnetic
field influencing the evolution of ${\bf R}(t)$.
Perhaps the anholonomy given by Eq.\ref{central}
will contribute to an intuitive understanding of
this {\it geometric magnetism}.
Incidentally, the restriction in this Letter to generalized
canonical families has a simple interpretation
in the context of Ref.\cite{br}:
for these families, the friction-like reaction
force ({\it deterministic friction}) is identically
zero\cite{robbins}.

The problem of generalizing the analysis of
this Letter to systems for which
Eq.\ref{soln} does not converge remains open,
and is related to the question of
whether or not ${\bf V}^c$ is closed
($\nabla\cdot{\bf V}^c=0$) in the general
case\cite{rb,robbins}.

In the time since the original submission of this
Letter, it has come to my attention that the classical
parallel transport studied herein has
been discussed by Montgomery\cite{mont} and
Golin et. al.\cite{golin}
(These authors obtain this transport
classically, rather than as the limit of
the quantal version.)
Eq.\ref{central}, however, is a new result.

\section*{ACKNOWLEDGMENTS}

I would like to thank W\l adek
\'Swi\c atecki and Jim Morehead for
stimulating discussions.
This work was supported by the Director, Office
of Energy Research, Division of Nuclear Physics
of the Office of High Energy and Nuclear Physics
of the U.S. Department of Energy under Contract No.
DE-AC03-76SF00098; and by the Department of Energy
under Grant No. DE-FG06-90ER40561.

\end{document}